\documentclass[prd,preprint,a4paper,superscriptaddress,nofootinbib,12pt]{revtex4}
\usepackage{epsfig}
\usepackage{amsmath}
\usepackage{amssymb}
\usepackage{amsthm}
\usepackage{stmaryrd}
\usepackage{url}
\usepackage{longtable}
\usepackage{mathrsfs}
\usepackage{amsfonts}
\usepackage{txfonts}
\usepackage[sans]{dsfont}
\def\fs{\footnotesize}
\def\frac#1#2{{#1\over#2}}
\def\bea{\begin{eqnarray}}
\def\eea{\end{eqnarray}}
\def\be{\begin{equation}}
\def\ee{\end{equation}}
\def\M{\mathcal{M}}
\def\P{\mathcal{P}}
\def\N{\mathcal{N}}

\begin{document}

\title{{\Large Classical basis for $\kappa$ -Poincar\'{e} Hopf algebra  and doubly special relativity theories}}
\author{A. Borowiec}
\email{borow@ift.uni.wroc.pl, borovec@theor.jinr.ru}
\affiliation{{\fs
Institute for Theoretical Physics, University of Wroclaw,
pl. Maxa Borna 9, 50-205 Wroclaw, Poland}}
\affiliation{{\fs Bogoliubov Laboratory of Theoretical Physics,
Joint Institute for Nuclear Research, Dubna,
Moscow region 141980, Russia}}
\author{A. Pacho{\l}$^1$}
\email{anna.pachol@ift.uni.wroc.pl}
\begin{abstract}

Several issues concerning quantum $\kappa-$Poincar\'{e} algebra are discussed and reconsidered here. We propose two different formulations of $\kappa-$Poincar\'{e}  quantum algebra.
Firstly we present a complete Hopf algebra formulae of $\kappa-$Poincar\'{e} in classical Poincar\'{e} basis. Further by adding one extra generator, which modifies the classical structure of Poincar\'{e} algebra, we eliminate non polynomial functions in the $\kappa-$ parameter.
Hilbert space representations of such algebras make Doubly Special Relativity (DSR) similar to the Stueckelberg's version of (proper-time) relativistic Quantum Mechanics.

\end{abstract}
\maketitle

\section{Introduction}

The concept of a quantum group was introduced more than 20 years ago in Refs.
\cite{Drin}- \cite{Fadeev} (see also \cite{Chiari}, \cite{Klimyk}), and since then the subject has been widely
investigated in different approaches and has gained in popularity and all sorts of applications
(see, e.g., \cite{Luk1}-\cite{BP}). One of the applications is to consider the notion of a quantum group as
noncommutative generalization of a symmetry group of the physical system,
which means that quantum group takes the place of symmetry group of
spacetime, i.e. Poincar\'{e} group.
Roughly speaking, quantum groups are the deformations of some classical structures as groups or Lie algebras,
which are made in the category of Hopf algebras. Similarly, quantum spaces are noncommutative generalizations (deformations) of ordinary spaces.
The most important in physics and mathematically the simplest one seem to be the canonical and the Lie-algebraic quantum
deformations. It has not taken long time to notice that in
description of the short-distance structure of spacetime (at the Planck
scale) the existing symmetries may be modified including deformation of
Poincar\'{e} symmetry. Moreover, it has been suggested that the symmetries of
the $\kappa$ -deformed Minkowski space should be described in terms of the
Hopf algebra \cite{Luk1}-\cite{LRZ}. The studies on this type of deformations
were inspired by \cite{Luk1}, where the quantum $\kappa$
-Poincar\'{e} algebra with the masslike deformation parameter was first
proposed.
The algebraic structure of the $\kappa$ -Poincar\'{e}
algebra has gained even more attention and since then has been intensively analyzed from
mathematical and physical point of view. For historical review see, e.g.,
\cite{Meljanac0702215}-\cite{BP} and the references therein.

A chance for physical application of this theory appeared when an
extension of special relativity was proposed in Refs. \cite{AC2,BACKG}, and another one, showing different point of view, in Refs. \cite{Smolin}. This extension
includes two observer-independent scales, the velocity of light and the
scale of mass, now called "Doubly Special Relativity" (DSR). Also, various phenomenological aspects of DSR theories have been studied in, e.g., \cite{phenom}. For comparison of these two approaches see, e.g., \cite{comparison}.  The connection
between $\kappa$ -deformation and DSR theory in first formulation (DSR1)
has been shown (see, e.g., \cite{BACKG}, \cite{GNbazy}, \cite{LukDSR}) including the conclusion that the
spacetime of DSR must be noncommutative as the result of Hopf structure of
this algebra.

The $\kappa$ -Poincar\'{e} algebra as well as DSR have been studied
extensively and have found many applications besides physics at the Planck
scale - gravity also in elementary particle physics and quantum field theory
(see, e.g., \cite{AC2}-\cite{Zegparticles} and references therein).
$\kappa$ -Poincar\'{e} Hopf algebra has been discovered in the so-called
standard basis \cite{Luk1} inherited from the anty-de Sitter basis by the contraction procedure. For this basis only the rotational sector remains algebraically undeformed. Introducing  bicrossproduct basis allows to leave the lorentzian generators undeformed.  This basis is the easier form of the $\kappa$ -Poincar\'{e} algebra basis and was postulated in \cite{MR}, \cite{LRZ}. In this form, the Lorentz subalgebra of the $\kappa$ -Poincar\'{e} algebra, generated by rotations and boosts is not deformed and the difference is only in the way the boosts act on momenta. There is also a change in co-algebraic sector, the coproducts are no longer trivial, which has the already mentioned consequence: the spacetime of DSR is noncommutative.

It is well known that the Drinfeld-Jimbo quantization algorithm relies on  simultaneous deformations of the algebraic and co-algebraic sectors and it is applicable to semisimple Lie algebras \cite{Drin,Jimbo}. In particular, this implies the existence of classical basis for Drinfeld-Jimbo quantized algebras. Strictly speaking, the Drinfeld-Jimbo technique cannot be applied to the Poincar\'{e}  non-semisimple algebra which has been obtained by the contraction procedure from the Drinfeld-Jimbo deformation of the anti-deSitter (simple) Lie algebra $\mathfrak{so}(3,2)$. Nevertheless, $\kappa-$Poincar\'{e} quantum group shares many properties of the original Drinfeld-Jimbo quantization. These include existence of the classical basis, the square of the antipode and the solution to the specialization problem (see Sect.III).

In the paper we define $\kappa$ -Poincar\'{e} (Hopf) algebra in its classical Poincar\'{e} Lie algebra basis. The constructions of such basis were previously investigated in several papers \cite{KosLuk}, \cite{Kos}.
Particularly, the explicit formulas expressing classical basis in terms of bicrossproduct one have been obtained therein~.
%have been suggested
%in several papers \cite{Kos}.
%However, according to best our knowledge,
Explicit formulas for coproducts can be found in different (realization dependent) context in  \cite{GNbazy}, \cite{Group21}, see also \cite{Meljanac0702215}.
To the best of our knowledge, there are no other examples of Drinfeld-Jimbo type deformation expressed in a classical Lie-algebraic basis.

It is known that different DSR models are defined by different choices of basis in the universal envelop of the Poincar\'{e} Lie algebra. They lead to different $\kappa-$Poincar\'{e} coproducts. Now we are in the position to demonstrate that when these models are compared in the same classical basis, they do differ by different operator realizations in the space of scalar-valued functions on a spacetime manifold.
This result in some sense allows us to distinguish between the description of DSR1 and DSR2 theories.
For the special choice of realization, we recover a well known bicrossproduct-form of $\kappa$
-Poincar\'{e} algebra and the standard DSR model. Moreover, according to the formalism developed in our previous paper \cite{BP}, we have a wide range of models and deformed dispersion relations related with them to our disposal (Sect.IV).

\section{$\protect\kappa$-Poincar\'{e} Hopf algebra in classical basis}

We shall use a standard so-called "physical" basis $(\mathcal{M}_{k},\mathcal{N}%
_{k},\mathcal{P}_{\mu })$ of the Poincar\'{e} Lie algebra $\mathfrak{P}^{1,3}
$ consisting of the Lorentz subalgebra $\mathfrak{L}^{1,3}$ of rotation $%
\mathcal{M}_{i}$  and boost $\mathcal{N}_i$ generators:
\begin{equation}  \label{L1}
\lbrack \mathcal{M}_{i},\,\mathcal{M}_{j}]\ =\ \imath \,\epsilon _{ijk}\,%
\mathcal{M}_{k}~,\qquad \lbrack \mathcal{M}_{i},\,\mathcal{N}_{j}]\ =\
\imath \,\epsilon _{ijk}\,\mathcal{N}_{k}~,\qquad \lbrack \mathcal{N}_{i},\,%
\mathcal{N}_{j}]\ =\ -\imath \,\epsilon _{ijk}\,\mathcal{M}_{k}~
\end{equation}
supplemented by Abelian four-momenta $\mathcal{P}_{\mu }=(\mathcal{P}_0, \mathcal{P}_k)$ ($%
\mu =0,\dots ,3\ ,k=1,2,3$) with the following commutation relations:
\begin{equation}  \label{L2}
\lbrack \mathcal{M}_{j},\,\mathcal{P}_{k}]=\imath \,\epsilon _{jkl}\,%
\mathcal{P}_{l}~,\qquad \lbrack \mathcal{M}_{j},\,\mathcal{P}%
_{0}]\;=\;0,\qquad \lbrack \mathcal{N}_{j},\,\mathcal{P}_{k}]\!\!=\!\!-%
\imath \,\delta _{jk}\,\mathcal{P}_{0}~,\quad \;\;[\mathcal{N}_{j},\,%
\mathcal{P}_{0}]\;=\;-\imath \,\mathcal{P}_{j}
\end{equation}
We take Lorentzian metric $\eta_{\mu\nu}=diag(-,+,+,+)$
for rising and lowering indices.

The algebra $(\mathcal{M}_{k},\mathcal{N}_{k},\mathcal{P}_{\mu })$ can be
extended in the standard way to a Hopf algebra by defining on the universal
enveloping algebra $U_{\mathfrak{P}^{1,3}}$ the coproduct $\Delta_0$, the counit
$\epsilon$, and the antipode $S_0$, where the nondeformed - primitive coproduct, the
antipode and the counit are given \be\label{undef} \Delta_0(X)=X\otimes
1+1\otimes X , \quad S_0(X)=-X , \quad \epsilon(X)=0 \ee
for $X\in \mathfrak{P}^{1,3}$. In addition $\Delta_0(1)=1\otimes 1$, $%
S_0(1)=1$ and $\epsilon(1)=1$. For the purpose of deformation one has to extend
further this Hopf algebra by considering formal power series in $\kappa^{-1}$, and correspondingly considering the Hopf algebra ($U_{\mathfrak{P}^{1,3}}[[\kappa^{-1}]],\cdotp%
,\Delta_0,S_0,\epsilon)$ as a topological Hopf algebra with the so-called "h-adic" topology \cite{Chiari,Klimyk}. Quantum
deformations of this Hopf algebra are controlled by classical $r$ -matrices satisfying the
classical Yang-Baxter (YB) equation: homogeneous or inhomogeneous.
The relation between classical $%
r$ -matrix $\mathfrak{r}$ and a universal (quantum) $r$ - matrix $\mathcal{R}$ reads as:
\begin{equation}  \label{qR}
\mathcal{R}=1+{1\over\kappa}\, \mathfrak{r}\, \ \
mod(\frac{1}{\kappa^2})
\end{equation}
where ${1\over\kappa}$ denotes the deformation parameter.
In the case of $r$ -matrices satisfying homogeneous YB equations the co-algebraic sector is
twist-deformed while algebraic one remains classical \cite{Drin}.
Additionally, one can also apply existing twist tensors to related Hopf module-algebras
in order to obtain quantized, e.g., spacetimes (see \cite{Meljanac0702215}-\cite{BP}).
In contrast Drinfeld-Jimbo quantization, corresponding to inhomogeneous $r-$ matrices, relies  on suitable deformation of the algebraic and co-algebraic sectors simultaneously.
Therefore, the classification of quantum deformations
is done by means of classification of the corresponding classical $r$ -matrices: homogeneous and inhomogeneous one.

In the case of relativistic  symmetries, such classification (complete for the Lorentz and almost-complete for Poincar\'{e} algebras) has been performed in Ref. \cite{Zakrzewski} (see also \cite{Lya} where this classification scheme has been extended). Particularly, $r$ -matrix which corresponds to $\kappa$ -deformation of Poincar\'{e} algebra is given by
 \be r=%\frac{1}{\kappa}
 N_i\wedge P^i \ee
and  it satisfies the inhomogeneous (modified) Yang-Baxter equation:
\be  [[r,r]]=%\frac{1}{\kappa^2}
M_{\mu\nu}\wedge P^\mu\wedge P^\nu \ee
Therefore, one does not expect to obtain the $\kappa$ Poincar\'{e} coproduct by twist.
However, most of the items on that list contain homogeneous $r-$matrices. Explicit twists for them have been provided in Ref. \cite{Varna} (for superization see \cite{VNT1,VNT2}); the corresponding quantization has been carried out in \cite{BLT2}.

Our purpose in this note is to formulate $\kappa-$Poincar\'{e} Hopf algebra in classical Poincar\'{e} basis. We would like to mention that complete treatment
of this problem was not considered before (see \cite{KosLuk}, \cite{Kos}).
One defines the deformed (quantized) coproducts $\Delta _{\kappa }$ and the antipodes $%
S_{\kappa }$ on $
\mathcal{U}\equiv U_{\mathfrak{P}^{1,3}}[[\kappa^{-1}]]$ leaving algebraic sector classical (untouched) like in the case of twisted deformation.
\begin{equation}
\ \Delta _{\kappa }\left( \mathcal{M}_{i}\right) =\Delta _{0}\left( \mathcal{%
M}_{i}\right)  \label{copM}
\end{equation}%
\begin{equation}
\ \Delta _{\kappa }\left( \mathcal{N}_{i}\right) =\mathcal{N}_{i}\otimes
1+\Pi_0^{-1}\otimes \mathcal{N}_{i}-\frac{1}{\kappa }\epsilon _{ijm}\mathcal{%
P}_{j}\Pi_0^{-1}\otimes \mathcal{M}_{m}  \label{copN}
\end{equation}%
\begin{equation}
\ \Delta _{\kappa }\left( \mathcal{P}_{i}\right) =\mathcal{P}_{i}\otimes
\Pi_0+1\otimes \mathcal{P}_{i}  \label{copPi}
\end{equation}%
\begin{equation}
\ \Delta _{\kappa }\left( \mathcal{P}_{0}\right) =\mathcal{P}_{0}\otimes
\Pi_0+\Pi_0^{-1}\otimes \mathcal{P}_{0}+\frac{1}{\kappa }\mathcal{P}%
_{m}\Pi_0^{-1}\otimes \mathcal{P}^{m}  \label{copP0}
\end{equation}%
and the antipodes
\begin{equation}  \label{sP1}
S_{\kappa }(\mathcal{M}_{i})=-\mathcal{M}_{i},\qquad S_\kappa(\mathcal{N}%
_{i})=-\Pi_0\mathcal{N}_{i}-\frac{1}{\kappa }\epsilon _{ijm}\mathcal{P}_{j}%
\mathcal{M}_{m}
\end{equation}%
\begin{equation}  \label{sP2}
S_\kappa(\mathcal{P}_{i})=-\mathcal{P}_{i}\Pi_0^{-1},\qquad S_\kappa(\mathcal{P}%
_{0})=-\mathcal{P}_{0}+\frac{1}{\kappa }\vec{\mathcal{P}}^{2}\Pi_0^{-1}
\end{equation}%
where
\begin{equation}  \label{Pi1}
\Pi_0\doteq\frac{1}{\kappa }\mathcal{P}_{0}+\sqrt{1-\frac{1}{\kappa ^{2}}%
\mathcal{P}^{2}}
%=\frac{1}{\kappa }\left( \mathcal{P}_{0}+\sqrt{\kappa ^{2}-%
%\mathcal{P}^{2}}\right)
%\end{equation}%
\quad \mbox{and}\quad
%\begin{equation}  \label{Pi2}
\Pi_0^{-1}\doteq\frac{\sqrt{1-\frac{1}{\kappa ^{2}}\mathcal{P}^{2}}-\frac{1}{%
\kappa }\mathcal{P}_{0}}{1-\frac{1}{\kappa ^{2}}\vec{\mathcal{P}}^{2}}%
%=\kappa \frac{\sqrt{\kappa ^{2}-\mathcal{P}^{2}}-\mathcal{P}_{0}}{\kappa
%^{2}-\vec{\mathcal{P}}^{2}}
\end{equation}%
are just shortcuts; $\mathcal{P}^{2}\doteq\mathcal{P}_{\mu }\mathcal{P}^{\mu
}\equiv\vec{\mathcal{P}}^{2}-\P _0^2$, and $\vec{\mathcal{P}}^{2}=\mathcal{P}%
_{i}\mathcal{P}^{i}$. Let us stress the point that above expressions are  formal
 power series in the parameter ${1\over\kappa}$
, e.g.,
 \be\label{sqrt}
 \sqrt{1-\frac{1}{\kappa ^{2}}\mathcal{P}^{2}}=\sum_{n\geq 0}\frac{(-1)^n}{\kappa^{2n}}\,\binom{0.5}{n}
 \,[\mathcal{P}^{2}]^n
 \ee
 where $\binom{0.5}{n}=\frac{0.5(0.5-1)\ldots (0.5-n+1)}{n!}$ are binomial coefficients. From the above one calculates
\begin{equation}  \label{Pi3}
\Delta_\kappa (\Pi_0)=\Pi_0\otimes \Pi_0,\quad \Delta_\kappa
(\Pi_0^{-1})=\Pi_0^{-1}\otimes \Pi_0^{-1}, \qquad S_\kappa(\Pi_0)=\Pi_0^{-1}
\end{equation}
as well as
\begin{equation}
\ \Delta _{\kappa }\left(\sqrt{1-\frac{1}{\kappa ^{2}}\mathcal{P}^{2}}\right) =\sqrt{%
1-\frac{1}{\kappa ^{2}}\mathcal{P}^{2}}\otimes \Pi_0 -\frac{1}{\kappa }%
\Pi_0^{-1}\otimes \mathcal{P}_{0}-\frac{1}{\kappa^2 }\mathcal{P}%
_{m}\Pi_0^{-1}\otimes\mathcal{P}^{m}
\end{equation}%
To complete the definition one leaves the counit $\epsilon$ undeformed. Let us observe that $\epsilon(\Pi_0)=\epsilon(\Pi_0^{-1})=1$. It is also worth noticing that the square of the antipode (\ref{sP1})-(\ref{sP2}) is given by a similarity transformation \footnote{In the case of twisted deformation the antipode itself is given by the similarity transformation.}, i.e. $$S_\kappa^2(X)=\Pi_0X\Pi_0^{-1}$$ .

Substituting now\be%\label{}
P_0\doteq\kappa\ln\Pi_0 , \qquad P_i\doteq \P _i\Pi_0^{-1}\quad \Rightarrow
\quad\Pi_0=e^{\frac{P_0}{\kappa}} \ee
one gets the deformed coproducts of the form
\begin{eqnarray}
\Delta_\kappa \left( P_{0}\right) &=&1\otimes P_{0}+P_{0}\otimes 1,\qquad
\Delta_\kappa \left( P_{k}\right) =e^{-\frac{P_{0}}{\kappa }}\otimes
P_{k}+P_{k}\otimes 1  \label{kP1} \\
\Delta_\kappa \left( \N_{i}\right) &=&\N_{i}\otimes 1+e^{-\frac{P_{0}}{%
\kappa }}\otimes \N_{i}-\frac{1}{\kappa }\epsilon _{ijm}P_{j}\otimes \N_{m}
\label{kP2}
\end{eqnarray}%
Similarly the commutators of new generators can be obtained as
\begin{equation}
\left[ \N_{i}, P_{j}\right] =-\frac{\imath}{2} \delta
_{ij}\left(\kappa\left( 1-e^{-\frac{2P_{0}}{\kappa }}\right) +\frac{1}{%
\kappa }\vec{P}^{2}\right) + \frac{\imath}{\kappa }P_{i}P_{j}
\end{equation}%
with the remaining one being the same as for Poincar\'{e} Lie
algebra (\ref{L1})-(\ref{L2}). This proves that our deformed Hopf algebra (%
\ref{L1})-(\ref{L2}), (\ref{copM})-(\ref{copP0}) is Hopf isomorphic to the
$\kappa $ -Poincar\'{e} Hopf algebra \cite{Luk1} written in its bicrossproduct
basis $(\M_i, \N_i, P_\mu)$ \cite{MR}. From now on we shall denote the Hopf algebra $U_{%
\mathfrak{P}^{1,3}}[[\kappa^{-1}]]$ (with $\kappa-$ deformed coproduct) by $%
\mathcal{U}(1,3)[[\kappa^{-1}]]$.\newline

\noindent The following immediate comments are now in order:\newline
i) Substituting $\N_i=M_{0i}$ and $\epsilon_{ijk}\M_k=M_{ij}$ the above
result easily generalizes to the case of \\
$\kappa $ -Poincar\'{e} Hopf
algebra in an arbitrary spacetime dimension $n$ (with the Lorentzian signature).
%Euclidean version of this Hopf algebra can be found in \cite{Meljanac0702215}%
\newline
ii) Although $\mathfrak{P}^{1,n-1}$ is Lie subalgebra of $\mathfrak{P}^{1,n}
$ the corresponding Hopf algebra $\mathcal{U}(1,n-1)[[\kappa^{-1}]]$ (with $%
\kappa-$deformed coproduct) is not Hopf subalgebra of $\mathcal{U}%
(1,n)[[\kappa^{-1}]]$.\newline
iii) Changing the generators by a similarity transformation $X\rightarrow
SXS^{-1}$ for $X\in (\M_i, \N_i, \P _\mu)$ leaves the algebraic sector (\ref%
{L1})-(\ref{L2}) unchanged but in general it changes coproducts (\ref{copM}%
)-(\ref{copP0}). Here $S$ is assumed to be an invertible element in $%
\mathcal{U}(1,3)[[\kappa^{-1}]]$. The both commutators and coproducts (\ref%
{copM})-(\ref{copP0}) are preserved provided that $S$ is group-like, i.e. $%
\Delta_\kappa (S)=S\otimes S$, e.g., $\Pi_0$. For physical applications it
might be also useful to consider another (nonlinear) changes of basis, e.g.
in the translational sector. Therefore the algebra $\mathcal{U}%
(1,3)[[\kappa^{-1}]]$ is a convenient playground for developing
Magueijo-Smolin type DSR theories \cite{Smolin}, \cite{Gosh} (DSR2) even if
we do not intend to take into account coproducts. But the coproducts are
there and can be used, e.g., in order to introduce an additional law for
four-momenta. In this situation the $\kappa$-deformed coproducts are not
necessary the privileged one and the additional law can be determined by,
e.g., the twisted coproducts \cite{BLT2}.
However $\kappa$-deformed coproducts are consistent with  $\kappa$-Minkowski commutation relations and give  $\kappa$-Minkowski spacetime module algebra structure \cite{Meljanac0702215}, \cite{BP}, \cite{BPprep}.

\section{New algebraic form of $\protect\kappa$-Poincar\'{e} Hopf algebra}

The mathematical formalism of quantum groups requires us to deal with formal
power series. Therefore the parameter $\kappa$ has to stay formal, i.e.,
undetermined. Particularly, we can not assign any particular numerical value
to it and consequently any fundamental constant of nature, like, e.g., the
Planck mass cannot be related with it. There are in principle two methods to
remedy this situation and allow $\kappa$ to admit constant value.

The first one is to reformulate algebra in such a way that all infinite
series will be eliminated on the abstract level. In the traditional Drinfeld-Jimbo
approach this is always possible  by using the so-called specialization method
or q-deformation (see e.g. \cite{Chiari}, \cite{Klimyk})
\footnote{In some physically motivated papers a phrase "q-deformation" is considered as an equivalent of Drinfeld-Jimbo deformation. In this section we shall, following general terminology of \cite{Chiari,Klimyk}, distinguish between "h-adic" and "q-analog" Drinfeld-Jimbo deformations  since they are not isomorphic.}
. The idea is to replace $%
\P _0$ by two group-like elements $\Pi_0, \Pi_0^{-1}$ and "forget" relation (\ref{Pi1}).
This provides,
for any specific (complex) numerical value $\kappa\neq 0$, a new
quantum algebra $\mathcal{U}_\kappa(1, 3)$. It is defined as a universal, unital
associative algebra generated by eleven generators $(\M_i, \N_i, \P _i,
\Pi_0, \Pi_0^{-1})$ being a subject of the standard Poincar\'{e} Lie-algebra
commutation relations (\ref{L1})-(\ref{L2}) except those containing $\P_0$.
These last should be replaced by the following new ones ($%
\Pi_0$ and $\Pi_0^{-1}$ are considered mutually inverse) \bea
%\begin{eqnarray}
[\P _i, \Pi_0]=[\M_j, \Pi_0]=0 ,\quad [\N_i, \Pi_0]=-\frac{\imath}{\kappa}\P %
_i \label{nk1}\\[7pt]
[\N_i, \P_j]=-\frac{\imath}{2}\delta_{ij}\left(\kappa(\Pi_0-\Pi_0^{-1})+\frac{1}{%
\kappa}\vec{\P }^2 \Pi_0^{-1}\right) \eea%\end{eqnarray}
The Hopf algebra structure is determined on $\mathcal{U}_\kappa(1, 3)$ by
the same formulae (\ref{copM})-(\ref{Pi3}) except those containing $\P _0$.
It should be noticed that these formulas contain only finite powers of the numerical parameter $\kappa$.
The generator $\P _0$ can be  now introduced as
\be
\P _0\equiv\P_0(\kappa)\doteq%\lim_{\kappa\rightarrow\infty}\frac{\kappa}{2}(\Pi_0-\Pi_0^{-1})=
\frac{\kappa}{2}\left(\Pi_0-\Pi_0^{-1}(1-\frac{1}{\kappa^2}\vec{\P }%
^2)\right) \ee
Thus subalgebra generated by elements $(\M_i, \N_i, \P _i,\P_0)$ is, of course, isomorphic to
the universal envelope of the Poincar\'{e} Lie algebra, i.e. $U_{\mathfrak{P}^{1,3}}\subset\mathcal{U}_\kappa(1, 3)$. But this is not a Hopf subalgebra.
Therefore, the original (classical) Casimir element $\mathcal{C}\equiv-\mathcal{P}^2
=\P _0^2-\vec{\P }^2$ has, in terms of  the generators $(\Pi_0, \Pi_0^{-1}, \vec{\mathcal{P}})$,
rather complicated form.  We can adopt to our disposal a simpler (central) element instead:
\be
\mathcal{C}_\kappa\doteq\kappa^2(\Pi_0^{}+\Pi_0^{-1}-2)-\vec{\P }%
^2 \Pi_0^{-1} \ee
which one may make responsible for deformed dispersion relations \cite{BP}. For comparison see, e.g., \cite{Casimir}.
Both elements are related by
\begin{equation}
\mathcal{C} =\mathcal{C}_\kappa\left(1+\frac{1}{4\kappa^2}\mathcal{C}_\kappa\right) \quad\mbox{and}\quad
\sqrt{1+\frac{1}{\kappa^2}{\mathcal{C}}}=1+\frac{1}{2\kappa^2}\mathcal{C}_\kappa
\end{equation}

Finally, one should notice that Hopf algebras $\mathcal{U}_\kappa(1,3)$ are isomorphic Hopf algebras for different values of $\kappa$. This is so since rescaling $\P_i\mapsto \frac{1}{\kappa}\P_i$ makes
$\mathcal{U}_\kappa(1, 3)\cong \mathcal{U}_1(1, 3)$.
%However they are isomorphic as coalgebras.

\section{Hilbert space realizations}
The second method allowing to specify value of $\kappa$ relies on
representation theory. Let us consider representation of the Poincar\'{e}
Lie algebra in a Hilbert space $\mathfrak{h}$. This leads to embedding of
the entire enveloping algebra $U_{\mathfrak{P}^{1,3}}$ into the space $\mathscr{L}(%
\mathfrak{h})$ of linear operators over $\mathfrak{h}$. Thus some elements
from $\mathcal{U}(1,3)[[\kappa^{-1}]]$ after substituting certain numerical
value for $\kappa$ can be considered as operators acting on $\mathfrak{h}$.
Roughly speaking specialization appears via spectral theorem on the level of
Hilbert space realization. Thus, in fact, one deals with a representation
of $\mathcal{U}_\kappa(1,3)$ instead of $\mathcal{U}(1,3)[[\kappa^{-1}]]$.
As an illustrative example one may consider a Stueckelberg's proper-time Hilbert space of square integrable complex-valued (wave)
functions on $\mathds{R}^4$, i.e. $\mathfrak{h}=L^2(\mathds{R}^4, d^4x)$ \cite{Menski} (see \cite{LRZ} for different representation).
There are canonical commutation relations between (local) momentum and position
operators \be
\lbrack p_\mu, x^\nu]=-\imath\, \delta^\nu_{\mu} ,\qquad\quad [p_\mu,
p_\nu]=[x_\mu, x_\nu]=0 \ee
represented by standard multiplication and differentiation operators: $x^\mu$
and $p_\mu=-\imath \partial_\mu$. Representation of the Poincar\'{e} Lie
algebra in this Hilbert space can be chosen, for example, as: \bea\label{DSRb}
\M_{i}=\frac{1}{2}\epsilon _{ijm}(x_{j}p _{m}-x_{m}p_{j}) ,\quad \N_{i}={%
\frac{\kappa }{2}}x_{i}\,\left( e^{-2\frac{p_{0}}{\kappa }}-1\right) +
x_{0}p _{i}-x_{i}\,\triangle \,+\frac{1}{\kappa }\,x^{k}p_{k}\,p_{i}\eea%
\bea
\mathcal{P}_{i}=p_{i}e^{\frac{p_{0}}{\kappa }}, \qquad \mathcal{P}%
_{0}=\kappa \sinh (\frac{p_{0}}{\kappa })+\frac{1}{2\kappa }\vec{p}^{2}e^{%
\frac{p_{0}}{\kappa }} \label{P0} \eea
where $\triangle =-\vec{p}^2$ denotes the Laplace operator.
 Now all
operators in the above formulas are well defined for constant value of $\kappa$ as Hilbert space operators. Moreover, it turns out that operators $(\M_i, \N_i, p_\mu)$ constitute the bicrossproduct
basis. Therefore, dispersion relations expressed in canonical momenta $p_\mu$
are the standard DSR. \be
\mathcal{C}_\kappa=\kappa ^{2}(e^{-{\frac{1}{2}}\frac{p_{0}}{\kappa }}-e^{{%
\frac{1}{2}}\frac{p_{0}}{\kappa }})^{2}+\triangle {e}^{\frac{p_{0}}{\kappa }%
}\ee
\begin{equation}
m_{0}^{2}=[2\kappa
\sinh (\frac{p_{0}}{2\kappa })]^{2}-\vec{p}^{2}e^{\frac{p_{0}}{\kappa }}
\end{equation}
One can notice that boost generators (\ref{DSRb}) in this representation are not Hermitian, because of the last term.
However Hermitian representation of the $\kappa$ -Poincar\'{e} algebra can be
determined as: \bea
\M_{i}=\frac{1}{2}\epsilon _{ijm}(x_{j}p _{m}-x_{m}p_{j}) ,\quad \N_{i}={%
\frac{\kappa }{2}}x_{i}\,\left( e^{-\frac{p_{0}}{\kappa }}-e^{\frac{p_{0}}{%
\kappa }}\right) + x_{0}p _{i}e^{-\frac{p_{0}}{\kappa }}-x_{i}\,\triangle
\,e^{-\frac{p_{0}}{\kappa }} %+\frac{1}{\kappa }%\,x^{k}p_{k}\,p_{i}
\eea\bea
\mathcal{P}_{i}=p_{i}, \qquad \mathcal{P}_{0}=\kappa \sinh (\frac{p_{0}}{%
\kappa })+\frac{1}{2\kappa }\vec{p}^{2}e^{-\frac{p_{0}}{\kappa }} \label{P0}
\eea
and the dispersion relation is%
\begin{equation}
m_{0}^{2}=[2\kappa \sinh (\frac{p_{0}}{2\kappa })]^{2}-\vec{p}^{2}e^{-\frac{%
p_{0}}{\kappa }}
\end{equation}
In both  cases above the representation of the element $\Pi_0$ in the Hilbert space realization
is given by the same formula $\Pi_0=e^{\frac{P_0}{\kappa}}$ (cf. also (17)) (see \cite{LRZ}, \cite{0307038}).\\

In the minimal case, connected with Weyl- Poincar\'{e} algebra, in physical $n=4$  dimensions \cite{BP} the representation of the Poincar\'{e} algebra $%
(M_{i},N_{i},\P_{\mu })$ reads as \begin{equation}
M_{i}=-\frac{\imath }{2}\epsilon _{ijm}(x_{j}\partial _{m}-x_{m}\partial
_{j})
\end{equation}%
\begin{eqnarray}
N_{i} &=&-x_{i}\left[{\frac{p_{0}}{2}}(2+\frac{%
p_{0}}{\kappa })+\,\triangle\right] \,\left( 1+\frac{p_{0}}{\kappa }\right)
^{-1} -\imath
x_{0}\partial _{i}
\end{eqnarray}
The generator $\Pi_0$ has now the form $\Pi_0=1+\frac{p_{0}}{\kappa }$ and the deformed Casimir operator
\begin{equation}
\mathcal{C}_\kappa=\frac{p_{0}^{2}-\vec{p}^{2}}{ 1+\frac{p_{0}}{\kappa}}
\end{equation}
leads to following dispersion relation
\begin{equation}
m_{0}^{2}\left( 1+\frac{p_{0}}{\kappa }\right) =p_{0}^{2}-\vec{p}^{2}
\end{equation}
which is not deformed for (free) massless particles.\newline
We close this note with an open question concerning choice of "physical" Casimir operator leading to correct dispersion relation and its operator realization.

\subsection*{Conclusions}

In this letter we have introduced two different Hopf algebras of $\kappa-$ Poincar\'{e} as quantum deformation of the Drinfeld-Jimbo type.
The first one is related to "h-adic" topology which forces the parameter $\kappa$ to stay abstract and undetermined. All formulae for coproducts
have been written intrinsically in a classical Lie algebra basis which is very typical for twisted Drinfeld deformation technique.
As it has been already explained, in the Introduction, the existence of such classical basis also for the Drinfeld - Jimbo deformations is a direct consequence of  their formalism. However the explicit construction is highly nontrivial mathematical
problem and to the best our knowledge it was investigated mainly for the case of $\kappa-$Poincar{\'e} \cite{KosLuk}, \cite{Kos}.
Particularly, the formulas expressing classical basis in terms of bicrossproduct one have been obtained therein~.
%Moreover, explicit formulas for coproducts can be found in different (realization dependent) context in  \cite{GNbazy}, \cite{Group21}, see also \cite{Meljanac0702215}.

The second definition relies on reformulating the Hopf algebra structure in such a way that infinite series disappear: it provides the one-parameter family of mutually isomorphic Hopf algebras labeled by a numerical (complex in general) parameter $\kappa$. So, the particular value of $\kappa$ becomes irrelevant.
From the physical point of view  one is allowed to work in the system of natural (Planck) units with $\hbar=G=c=1$ without changing mathematical properties
of the underlaying quantum model. In this way the  so-called specialization problem for the deformation parameter $\kappa$ has been solved
\footnote{Similar problem in the bicrossproduct basis has been previously studied in \cite{Stachura}.}. Finally, it has been shown that
%the (proper-time) Hilbert space realizations of such Hopf algebras has been considered as well.
different (proper-time) Hilbert space representation of this algebra can  be understood as the corresponding to different
DSR type models providing different dispersion relations. Therefore, we believe that our research might be also helpful to distinguish
between two approaches to doubly special relativity theories.

\subsection*{Acknowledgements}
This paper has been supported by MNiSW Grant No. NN202 318534 and the
Bogliubov-Infeld Program. The authors acknowledge helpful discussions with S. Meljanac, J. Kowalski-Glikman and V.N. Tolstoy. Special thanks are due to J.
Lukierski for valuable comments and reading the manuscript.


\begin{thebibliography}{99}
\bibitem{Drin} V. Drinfeld, \textit{Proceedings of the International
Congress of Mathematicians}, Berkeley, 1986 (American Mathematical Society,
Providence, 1987); V. Drinfeld, \textit{Sov. Math. Dokl.} \textbf{32}, 254
(1985);
\bibitem{Jimbo} M. Jimbo, \textit{Lett. Math. Phys.} \textbf{10}, 63 (1985);
\bibitem{Woronowicz} S. L. Woronowicz, \textit{Comm. Math. Phys.} \textbf{111%
}, 613 (1987);
\bibitem{Fadeev} L.D. Fadeev, N.Yu. Reshetikhin, L.A. Takhtajan, \textit{%
Leningrad Math. Journ.} \textbf{1}, 193 (1990);
\bibitem{Chiari} V.~Chari and A.~Pressley, \textit{A Guide to Quantum Groups}%
, Cambridge University Press, (1994);
\bibitem{Klimyk}
A.Klimyk, K.Schmudgen "Quantum Groups and their representations" Springer-Verlag, Berlin Heidelberg 1997;
\bibitem{Luk1} J. Lukierski, A. Nowicki, H. Ruegg, V. N. Tolstoy, \textit{%
Phys. Lett.} \textbf{B264}, 331 (1991);
\bibitem{LNR} J. Lukierski, A. Nowicki, H. Ruegg, \textit{Phys. Lett.}
\textbf{B293}, 344 (1992);
\bibitem{Zakrz} S. Zakrzewski, J. Phys. A27, 2075 (1994);
\bibitem{MR} S. Majid, H. Ruegg, \textit{Phys. Lett.} \textbf{B334}, 348
(1994) [arXiv:hep-th/9405107];
\bibitem{LRZ} J. Lukierski, H. Ruegg, W. Zakrzewski, \textit{Ann. Phys.},
\textbf{243}, 90 (1995);
\bibitem{0307038} P. Kosiñski, P. Maœlanka, J. Lukierski, A. Sitarz, \textit{Proceedings of the Conference "Topics in Mathematical Physics, General Relativity and Cosmology"}, On the occasion of the 75th Birthday of Jerzy F. Plebanski, 17.09-20.09 2002, Mexico City, Eds: H. Garcia-Compean et.al., World Scientific (2003) [arXiv:hep-th/0307038];
\bibitem{BLT2} J. Lukierski, H. Ruegg, V.N. Tolstoy, A. Nowicki, \textit{J.
Phys.}, \textbf{A27}, 2389 (1994), [arXiv:hep-th/9312068];
A. Borowiec, J. Lukierski, V. N. Tolstoy,
\textit{Eur. Phys. J.}, \textbf{C57}, 601 (2008), %DOI 10.1140
[arXiv:0804.3305]; A. Borowiec, J. Lukierski, V.N. Tolstoy, \textit{Eur.
Phys. J.}, \textbf{C48}, 633 (2006), [arXiv:hep-th/0604146]; A. Borowiec, J.
Lukierski, V. N. Tolstoy, \textit{Eur. Phys. J.}, \textbf{C44}, 139 (2005),
[arXiv:hep-th/0412131]; A. Borowiec, J. Lukierski, V. N. Tolstoy, \textit{%
Czech. J. Phys}, \textbf{55}, 11 (2005), [arXiv:hep-th/0510154]; A.
Borowiec, J. Lukierski, V.N. Tolstoy, \textit{Mod. Phys. Lett.}, \textbf{A18}%
, 1157 (2003), [arXiv:hep-th/0604146];
\bibitem{Meljanac0702215} S. Meljanac, S. Kresi\'{c}-Juri\'{c}, M. Stoji\'{c}%
, \textit{Eur. Phys. J.} \textbf{C51},229 (2007), [arXiv:hep-th/0702215];
\bibitem{Melj}
T.R. Govindarajan, Kumar S. Gupta, E. Harikumar , S. Meljanac, D. Meljanac,
\textit{Phys.Rev.} {\bf D77}:105010 (2008), [arXiv:0802.1576];
S. Meljanac, S. Kresic-Juric, \textit{J.Phys.} {\bf A41}:235203 (2008), [arXiv:0804.3072];
N. Durov, S. Meljanac, A. Samsarov, Z. Skoda, \textit{J. Algebra},
{\bf 309}, 318 (2006), [arXiv:hep-th/0604096];
 S. Meljanac, M. Stoji\'{c}, \textit{Eur. Phys. J.} \textbf{C47}%
, 531 (2006), [arXiv:hep-th/0605133];
 S. Meljanac, A. Samsarov, M. Stoji\'{c}, K. S. Gupta,
\textit{Eur. Phys. J.} \textbf{C53}, 295 (2008), [arXiv:0705.2471];
\bibitem{BP} A. Borowiec, A. Pachol, \textit{Phys. Rev.} \textbf{D 79},
045012 (2009) [arXiv:0812.0576];
%\bibitem{AC1}
\bibitem{AC2} G. Amelino-Camelia, \textit{Int. J. Mod. Phys.} \textbf{D11},
35 (2002) [arXiv:gr-qc/0012051];
G. Amelino-Camelia, \textit{Phys. Lett.} \textbf{B510}, 255
(2001) [arXiv:hep-th/0012238];
\bibitem{BACKG} B. Bruno, G. Amelino-Camelia, J. Kowalski-Glikman, \textit{%
Phys. Lett.} \textbf{B522}, 133 (2001) [arXiv:hep-th/0107039]; J. Kowalski-Glikman, \textit{Phys. Lett.} \textbf{A286}, 391 (2001)
[arXiv:hep-th/0102098];
\bibitem{Smolin} J. Magueijo, L. Smolin, \textit{Phys.Rev.Lett.} \textbf{88}%
:190403 (2002), [arXiv:hep-th/0112090]; J. Magueijo, L. Smolin, \textit{%
Phys.Rev.} \textbf{D67}:044017 (2003), [arXiv:gr-qc/0207085];
\bibitem{phenom} G. Amelino-Camelia, L.Smolin and Starodubtsev,
{\it Class. Quant. Grav.} {\bf 21}, 3095 (2004) [arXiv:hep-th/0306134];
 G. Amelino-Camelia, J. Kowalski-Glikman, G. Mandacini and A. Pro-
caccini, {\it Int. J. Mod. Phys.} {\bf  A20}, 6007 (2005) [arXiv:gr-qc/0312124]
 C. Rovelli "A note on DSR" [arXiv:0808.3505];
\bibitem{comparison} G. Amelino-Camelia, D. Benedetti, F. D'Andrea, A. Procaccini  \textit{ Class. Quant. Grav.}  {\bf 20}, 5353 (2003) [arXiv:hep-th/0201245];
\bibitem{GNbazy} J. Kowalski-Glikman and S.Nowak, \textit{Phys. Lett.} {\bf B539}, 126 (2002)
[arXiv:hep-th/0203040];
 L. Freidel, J. Kowalski-Glikman, S. Nowak, \textit{Int. J. Mod. Phys.} \textbf{A23}, 2687 (2008) [arXiv:0706.3658];
\bibitem{LukDSR} J. Lukierski, Proceedings of the International Workshop "Supersymmetries and Quantum Symmetries" (SQS'03 July 2003), Dubna 2004, Eds. E. Ivanov, A. Pashnev, [arXiv:hep-th/0402117];
\bibitem{ACAD} G. Amelino-Camelia, M. Arzano, L. Doplicher, \textit{A
relativistic spacetime odyssey}, 497, Florence (2001) [arXiv:hep-th/0205047];
\bibitem{Zegparticles}
M. Daszkiewicz, J. Lukierski, M.Woronowicz, \textit{Mod. Phys. Lett.} \textbf{A23}, 653 (2008) [arXiv:hep-th/0703200];
 M. Daszkiewicz,  J. Lukierski, M. Woronowicz, \textit{Phys. Rev.} \textbf{D77}, 105007  (2008) [arXiv:hep-th/0708.1561]; M. Dimitrijevic, L. Jonke, L. Möller, E. Tsouchnika, J. Wess, M. Wohlgenannt
 {\it Eur. Phys. J.} {\bf C31}, 129 (2003) [arXiv:hep-th/0307149];
\bibitem{KosLuk} P. Kosinski, J. Lukierski, P. Maslanka, J. Sobczyk
 \textit{Mod. Phys. Lett.}  \textbf{A10} , 2599 (1995)  [arXiv:hep-th/9412114];
\bibitem{Kos}  P. Maslanka, \textit{J. Math. Phys.} \textbf{34}, 12 (1993);
P. Kosinski, P. Maslanka "The Kappa-Weyl group and its algebra" Proceedings of the XXII Max Born Symposium "Quantum,
Super and Twistors" (Wydawnictwo Uniwersytetu Wroclawskiego, Wroclaw, 2007)
[arXiv:q-alg/9512018] ;
P. Kosinski, P. Maslanka,  \textit{Phys.Rev.} {\bf D68}:067702 (2003), [arXiv:hep-th/0211057];
\bibitem{Group21} J. Lukierski "$\kappa$ -Deformations of Relativistic Symmetries: Some Recent Developments" Quantum Group Symposium at GROUP 21, eds. H.-D. Doebner and V.K. Dobrev, Heron Press, Sofia 1997; J. Lukierski, A. Nowicki \textit{Int.J.Mod.Phys.} \textbf{A18}, 7 (2003) [arXiv:hep-th/0203065];
\bibitem{Zakrzewski} S. Zakrzewski, \textit{Lett. Math. Phys.}, \textbf{32},
11 (1994); %\bibitem{Z1}
S. Zakrzewski, \textit{Commun. Math. Phys.}, \textbf{187}, 285 (1997),
[arXiv:q-alg/9602001];
\bibitem{Lya} V.D. Lyakhovsky, \textit{Reports Mat. Phys.}, \textbf{61}, 213 (2008), M. Daszkiewicz,
\textit{Rept. Mat. Phys.}, \textbf{63}, 263-277 (2009) [arXiv:0812.1613]
\bibitem{Varna} V.N. Tolstoy, \emph{Twisted Quantum Deformations of Lorentz
and Poinca\'re algebras} In: \textit{Lie Theory and Its Applications in
Physics. LT-7}. Proceedings of the VII International Workshop (Varna,
Bulgaria, 18-24 June, 2007). Eds: H.-D. Doebner, V.K. Dobrev. Publ: Heron
Press, Sofia, 441 (2008); V.N. Tolstoy, \emph{Quantum Deformations of
Relativistic Symmetries}, Proceedings of the XXII Max Born Symposium
''Quantum, Super and Twistors'', PWN, Warszawa, 2008, [arXiv:0704.0081];
\bibitem{VNT1} V.N. Tolstoy, \textit{Proc. of International Workshop
"Supersymmetries and Quantum Symmetries (SQS'03)", Russia, Dubna, July,
2003, eds: E. Ivanov and A. Pashnev, Publ. JINR, Dubna}, p. 242 (2004),
[arXiv:math/0402433];
\bibitem{VNT2} V.N. Tolstoy,
in: \textit{Nankai Tracts in Mathematics "Differential Geometry and
Physics". Proceedings of the 23-th International Conference of Differential
Geometric Methods in Theoretical Physics (Tianjin, China, 20-26 August,
2005). Editors: Mo-Lin Ge and Weiping Zhang. Wold Scientific}, 2006, Vol.
10, 443-452 [arXiv:math/0701079]; A. Borowiec, J. Lukierski, V. N. Tolstoy,
\emph{New twisted quantum deformations of D=4 super-Poincar\'{e} algebra},
Proceedings of International Workoshop: Supersymmetries and Quantum Symetries (SQS'07), Dubna, Russia 2008 ed S Fedoruk and E Ivanov pp 205–-15 [arXiv:0803.4167];
\bibitem{Gosh}
F. Girelli, E. R. Livine  \textit{Braz. J. Phys.}, \textbf{35}, 432 (2005)
[arXiv:gr-qc/0412079]; "Physics of Deformed Special Relativity: Relativity Principle revisited" [arXiv:gr-qc/0412004];
Subir Gosh, \textit{Phys.Rev.} \textbf{D75}:105021 (2007),
[arXiv:hep-th/0702159];
\bibitem{Casimir}
H. Ruegg, V. N. Tolstoy  \textit{Lett. Math. Phys.} \textbf{32}, 85 (1994);
A. Nowicki, E. Sorace, M. Tarlini, \textit{Phys. Lett.} \textbf{B302}, 419 (1993) [arXiv:hep-th/9212065];
\bibitem{Menski}
E. C. G. Stueckelberg \textit{Helv. Phys. Acta} \textbf{14}, 322 (1941);
 J.H. Cooke \textit{Phys. Rev.} \textbf{166}, 1293 (1968);
J.E. Johnson \textit{Phys. Rev.} \textbf{181}, 1755 (1969); \textit{Phys. Rev.} \textbf{D3}, 1735 (1971);
A.A. Broyles \textit{Phys. Rev.} \textbf{D1}, 979 (1970);
J.J. Aghassi, P. Roman, R.M.Santilli \textit{Phys. Rev.} \textbf{D1}, 2753 (1970);
M.B. Mensky, \textit{Commun.Math.Phys.} \textbf{47}, 97 (1976);
\bibitem{BPprep} A. Borowiec, A. Pacho{\l} 2010 {\it Parameter free formulation of ƒ$\kappa$-Minkowski spacetime and $\kappa$-Poincare Lie group},
Proc. 13th Int. Conf. on Symmetry Methods in Physics (SYMPHYS 2009) Yad. Fizika (at press)
\bibitem{Stachura} P. Stachura, \textit{Reports Math. Phys.} \textbf{57}, 233 (2006);
\end{thebibliography}
\end{document}